\documentclass[
preprint,
superscriptaddress,
amsmath,amssymb,
aps,
prl,
floatfix,
]{revtex4-2}

\usepackage{graphicx}           
\usepackage{dcolumn}            
\usepackage{bm}                 
\usepackage{hyperref}           

\usepackage{siunitx}
\usepackage{fp}
\newcommand{\AlGaAs}[1][0.929]{\FPeval{\result}{round(1-#1,3)}Al\textsubscript{#1}Ga\textsubscript{\result}As}
\newcommand{\AlGaAsx}{Al\textsubscript{x}Ga\textsubscript{1-x}As}
\usepackage[inline]{enumitem}

\begin{document}


\title{Precise Measurement of Refractive Indices in Thin Film Heterostructures}

\author{Lukas W. Perner}
    \email{lukas.perner@univie.ac.at}
    \affiliation{Christian Doppler Laboratory for Mid-IR Spectroscopy and Semiconductor Optics, Faculty Center for Nano Structure Research, Faculty of Physics, University of Vienna, Boltzmanngasse 5, 1090 Vienna, Austria.}
    \affiliation{Vienna Doctoral School in Physics, University of Vienna, Boltzmanngasse 5, 1090 Vienna, Austria.}
\author{Gar-Wing Truong}
    \affiliation{Thorlabs Crystalline Solutions, 114 E Haley St., Suite G, Santa Barbara, California 93101, USA}
\author{David Follman}
    \affiliation{Thorlabs Crystalline Solutions, 114 E Haley St., Suite G, Santa Barbara, California 93101, USA}
\author{Maximilian Prinz}
    \affiliation{Christian Doppler Laboratory for Mid-IR Spectroscopy and Semiconductor Optics, Faculty Center for Nano Structure Research, Faculty of Physics, University of Vienna, Boltzmanngasse 5, 1090 Vienna, Austria.}
\author{Georg Winkler}
    \affiliation{Christian Doppler Laboratory for Mid-IR Spectroscopy and Semiconductor Optics, Faculty Center for Nano Structure Research, Faculty of Physics, University of Vienna, Boltzmanngasse 5, 1090 Vienna, Austria.}
\author{Stephan Puchegger}
    \affiliation{Faculty Center for Nano Structure Research, Faculty of Physics, University of Vienna, Boltzmanngasse 5, 1090 Vienna, Austria.}
\author{Garrett D. Cole}
    \affiliation{Thorlabs Crystalline Solutions, 114 E Haley St., Suite G, Santa Barbara, California 93101, USA}
\author{Oliver H. Heckl}
    \affiliation{Christian Doppler Laboratory for Mid-IR Spectroscopy and Semiconductor Optics, Faculty Center for Nano Structure Research, Faculty of Physics, University of Vienna, Boltzmanngasse 5, 1090 Vienna, Austria.}

\date{\today}

\begin{abstract}
We present a robust, precise, and accurate method to simultaneously measure the refractive indices of two transparent materials within an interference coating. This is achieved by measuring both a photometrically accurate transmittance spectrum and the as-grown individual layer thicknesses of a thin-film multilayer structure. These measurements are used for a TMM-based curve-fitting routine which extracts the refractive indices and their measurement uncertainties via a Monte-Carlo-type error propagation. We  demonstrate the performance of this approach by experimentally measuring the refractive indices of both, GaAs and \AlGaAs{}, as present in an epitaxial distributed Bragg reflector. A variety of devices can be used to obtain the transmittance spectrum (e.g., FTIR, grating-based spectrophotometer) and layer thicknesses (e.g., SEM, TEM, AFM), the discussed approach is readily adaptable to virtually any wavelength region and many transparent material combinations of interest. The subsequent model-fitting approach yields refractive index values with $10^{-4}$-level uncertainty for both materials.

\end{abstract}

\maketitle

Obtaining accurate and precise values for the refractive index of transparent materials is paramount for the design and production of many optical devices, among them vertical-cavity surface-emitting lasers (VCSELs)~\cite{2001Vertical-CavityLasers}, light emitting and super-luminescent diodes~\cite{Lin1997ExtremelyDiodes}, photodetectors~\cite{Kozlowski1991LWIRArray}, photovoltaics~\cite{Cotal2009IIIVPhotovoltaics}, thin-film multilayer structures such as anti-reflection (AR) or high-reflectivity (HR) coatings and other spectral interference filters~\cite{Rempe1992MeasurementInterferometer}, as well as high-performance substrate-transferred optical interference coatings~\cite{Cole2013TenfoldCoatings,Cole2016,Winkler2021Mid-infraredPpm}.
In most of these applications, the choice of materials relies on refractive index values published in literature. Therefore, especially in less well-studied wavelength regions such as the mid-infrared (MIR), a plethora of applications rely on extrapolations of data originally obtained in the near-infrared (NIR) or via measurements taken for different sample types of the same material (e.g., bulk vs thin film and amorphous vs. crystalline samples).

While model extrapolations have their merit to generate general insights, inaccuracies lead to significant deviation from target parameters, such as design wavelength or reflectivity and, often costly, iterations in the production of high-end optical devices. In most cases, the required uncertainties can only be achieved by measuring the optical response of materials and fitting of an appropriate model in a restricted wavelength region of interest.

\begin{figure}
    \includegraphics{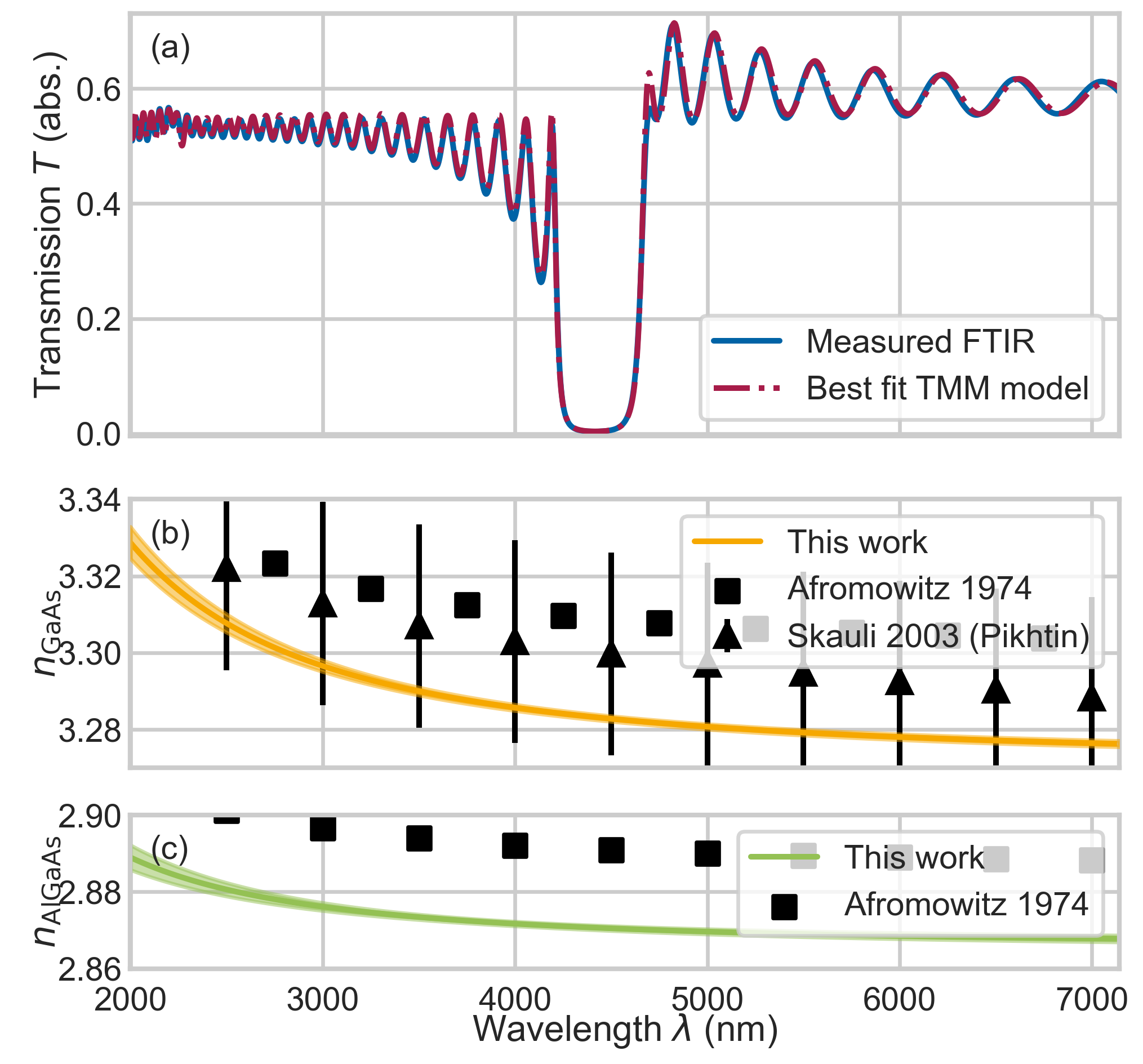}
    
    \caption{(a) Transmittance spectra of the measured sample multilayer structure as measured in FTIR (blue) and the best fit model as obtained from a nonlinear least-squares regression with the TMM model (red) based on the physical layer thicknesses shown in Fig.~\ref{fig:SEM_overview}(a). (b) Final results for the refractive index of GaAs. (c) Final results for the refractive index of \AlGaAs{}. In (b) and (c), error bands/bars are given as the fourfold standard uncertainty $4s$. For comparison, we include results from~\cite{Afromowitz1974,Skauli2003ImprovedOptics,Palmer2002Mid-infraredAlAs}.}
    \label{fig:T_spectrum_fit_and_n_results} 
\end{figure}

However, such thin-film structures are routinely produced with ultra-high purity in both crystalline and amorphous form via well-established deposition methods, such as molecular beam epitaxy (MBE)~\cite{Cho1971FilmTechniques, Pohl2020EpitaxySemiconductors} for crystalline thin films and ion beam sputtering (IBS) for amorphous films~\cite{Sites1983IonCoatings}.
Furthermore, multilayer structures can combine two (or more) materials in a multilayered system with very high purity, owing to extremely low levels of impurities.
This is demonstrated by the ultra-high reflectivity when used to produce distributed Bragg reflectors (DBR) at various wavelengths~\cite{Rempe1992MeasurementInterferometer, Cole2013TenfoldCoatings, Cole2016, Winkler2021Mid-infraredPpm, Truong2022Transmission-dominated000}.
Such DBRs are nominally designed as a periodic two-material structure with alternating high- and low-refractive-index layers of quarter-wave optical thickness at a certain target wavelength $\lambda_\mathrm{d}$. Due to this design, reflections at each layer interface interfere constructively, leading to a characteristic stopband-like transmittance spectrum (see. Fig.~\ref{fig:T_spectrum_fit_and_n_results}(a)).
In such a spectrum, both the depth and the width of the main feature around $\lambda_\mathrm{d}$ and its side lobes are a function only of the refractive indices of the DBR materials, for a given number and thickness of individual layers. Hence, the transmittance spectrum highly constrains the refractive index values of a model fitted to the data. The above-mentioned properties make thin-film structures, and especially DBR structures, a promising system to probe the refractive index of the constituent materials.

\begin{figure*}
    \includegraphics[width=\columnwidth]{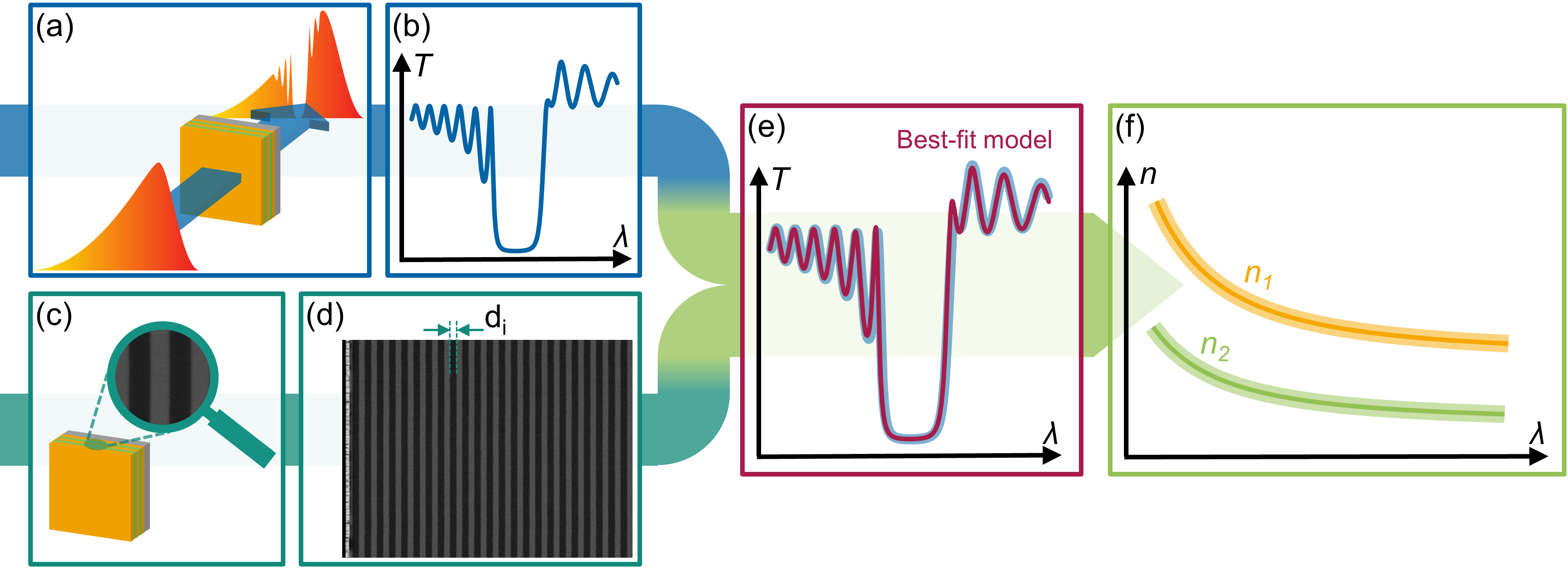}
    
    \caption{
    Schematic outlining the measurement method. It relies on two measurements on a thin film heterostructure: (a) First, A spectrometric transmittance (or reflectance, not depicted) measurement. (b) From this, we acquire the optical response of the sample. (c) Second, a cross-sectional SEM (or TEM/AFM, not depicted) micrograph. (d) From this, we extract the individual layer thicknesses $d_i$. (e) These two measurements are used for a TMM-based best-fit modeling approach. (f) Finally, from this fit, we infer both refractive indices, $n_1$ and $n_2$, as well as their respective uncertainties (using a Monte-Carlo-type propagation of the measurement uncertainties from (a)--(d)). The SEM micrograph depicted in (d) is one of a total of four of the sample used in the proof-of-principle experiment.}
    \label{fig:Intro}
\end{figure*}

\begin{figure}
    \includegraphics{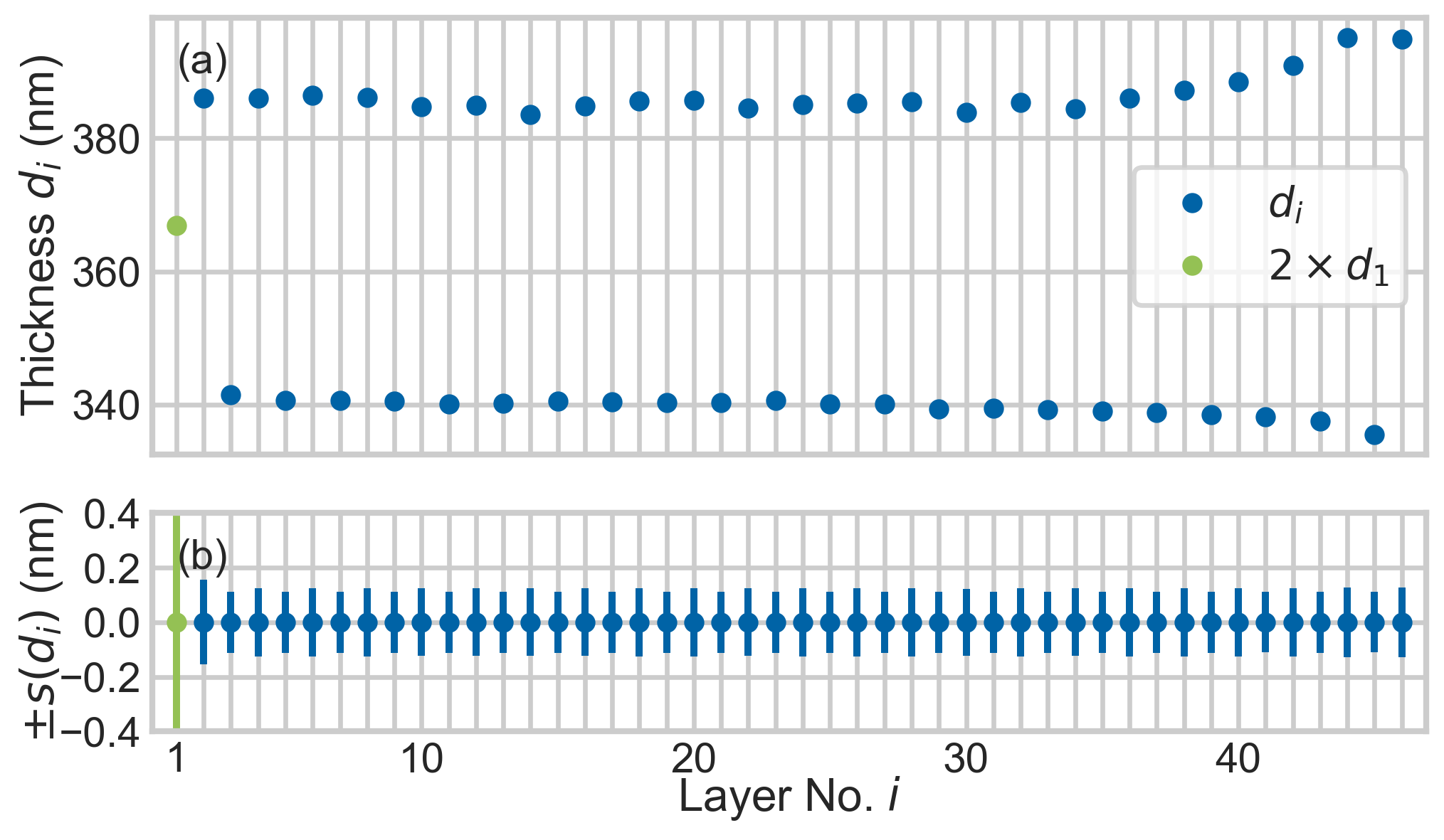}
    
    \caption{\label{fig:SEM_overview}
    (a) Measured layer thicknesses. Note that $d_1$ (green) was multiplied by a factor of 2. (b) Error bars showing $1s$ standard uncertainty for all mean values given in (a).}
\end{figure}

Compared to established methods probing bulk single-material samples, including high-resolution Fourier-transform infrared (FTIR) refractometry~\cite{Kaplan1998FourierRefractometry, Milosevic2013SpectroscopicIndex, Skauli2003ImprovedOptics} or prism-based methods~\cite{Onodera1983Refractive-indexMethod}, probing DBRs can avoid exposure to atmosphere for the materials under test, which prevents the formation of oxide layers in quickly-oxidizing materials, such as \AlGaAsx{}~\cite{Reinhardt1996OxidationAir}, thereby reducing systematic errors.
A recently-published study, using spectroscopic ellipsometry~\cite{Papatryfonos2021RefractiveRegion} to probe \AlGaAsx{}, addressed oxide layer formation by incorporating it in their modelling approach. This approach remains sensitive to the properties of the oxide layer, as only a single MBE-grown \AlGaAsx{} layer (with $0.097\leqslant x \leqslant 0.411$) on a GaAs substrate is probed in reflection. Furthermore, the approach requires a calibration step on pure GaAs and immersion of all samples in HCl prior to measurement, as well as accurate knowledge of the alloy composition via XRD and the material bandgaps via photoluminescence measurements.
Palmer et al. devised a purely optical approach~\cite{Palmer2002Mid-infraredAlAs}. It requires the measurement of the reflectance spectrum of the same DBR sample in two different wavelength regions, the region of interest, as well as one where the refractive indices of the materials is already known. In the region of known refractive index, a curve-fitting routine is used to obtain an estimate for the mean layer thickness per material. Subsequently, another best-fit model is calculated in the wavelength region of interest to obtain best-fit parameters for the refractive indices, with the previously obtained layer thicknesses now as fixed parameters. Since only a mean layer thickness is obtained in the first step, this approach assumes a perfectly periodic layer stack and does not take into account potential variations of individual layer thicknesses in an as-grown multilayer structure (as seen in Fig.~\ref{fig:SEM_overview}(a)).

Here, we propose a new technique for the precise determination of refractive indices.
As will be discussed in detail below, the proposed method involves three main steps:
\begin{enumerate*}[label=(\roman*)]
    \item measuring a photometrically-accurate (i.e., unaffected by systematic effects such as detector nonlinearity) transmittance (or reflectance) spectrum of a multilayer sample
    \item obtaining individual physical layer thicknesses from a cross-sectional measurement
    \item perform a best-fit routine using an appropriate model,
\end{enumerate*}
as outlined by the schematic in Fig.~\ref{fig:Intro} (a detailed report of each step is found in~\cite{PernerSimultaneously-MeasuredGaAs}).
This routine yields a propagated standard uncertainty on the $10^{-4}$ level for both materials.

Note that the measurement devices used to obtain individual layer thicknesses and the transmittance spectrum can be chosen depending on the desired level of uncertainty, measurement-specific parameters (e.g., wavelength range or material composition), and availability. However, the accuracy and precision of the measurements will ultimately drive the level of uncertainty that can be achieved for the refractive indices.

The first step is choosing an appropriate sample. As discussed above, a standard two-material quarter-wave DBR of moderate reflectivity (\SI{\sim 90}{\percent} at the stopband center), with a stopband feature in the wavelength region of interest is ideal. The multilayer structure can be deposited on a substrate by a variety of processes that yield homogeneous, highly-pure layers with abrupt surfaces. As the exact layer thicknesses will be measured in our process, a considerable amount of variation is tolerable as long as the transmittance spectrum still exhibits broadband features (much bigger than the resolution of the acquired transmittance spectrum). The substrate can be chosen according to the limitations of the fabrication process. However, it is beneficial if it consists of a well-characterized material or the same material as one of the thin-film layers. In general, crystalline samples will be easier to process for the necessary measurements and be of higher purity, whereas amorphous samples allow for more material combinations and are more cost-effective to obtain.

The second step for refractive index determination begins with taking a photometrically accurate transmittance spectrum at normal incidence.
The spectrometric device can be chosen based on specific requirements (wavelength range, accuracy) and availability. In general, FTIR devices are a good choice, as they are well established, widely available and, if care is taken, can obtain accurate broadband spectra.
With most spectrometers, it is easier to achieve photometric accuracy when the measured values are all in the $10^{-1}$ range, as systematic measurement uncertainties, such as nonlinearities of optical detectors~\cite{Theocharous2004AbsoluteRegion,Theocharous2008AbsoluteDetectors,Theocharous2008AbsoluteDetector,Theocharous2012AbsoluteInfrared}, become more manageable when avoiding measurements covering several orders of magnitude (although even then, excellent results can be obtained, as we show below).
In principle, the presented approach can also be used with a reflectance spectrum (instead of transmittance). However, without specialized equipment, achieving high photometric accuracy in reflectance measurements generally involves additional calibration steps, adding another source of uncertainty.

The third step is to determine the as-fabricated individual layer thicknesses. For that, the sample is split in half to expose the cross-sectional layer structure as close to the spot probed for transmittance as possible. We tested this for crystalline samples, where revealing the cross-section via cleaving is straightforward. However, cross-sectional images of comparable quality are obtainable by cutting and polishing an amorphous specimen.
Then, the sample is subjected to cross-sectional imaging. While, in principle, other measurement techniques, such as tunneling electron microscopy (TEM) or atomic force microscopy (AFM), can be used to retrieve the layer structure, we chose a scanning-electrom microscope (SEM). As detailed below, the SEM-based procedure allows for many independent measurements of the layers from a single cross-sectional image, as it can be evaluated line-by-line and averaged afterwards. This yields highly accurate measurements, effectively limited by the uncertainty associated with the SEM length calibration standard.

The final step combines the measured transmittance (or reflectance) spectra along with the measured physical layer thicknesses together in a model using the transmission matrix method (TMM)~\cite{Born1999PrinciplesOptics, Byrnes2021MultilayerCalculations, Luce2022TMM-FastTutorial}.
The TMM calculates the transmittance $T$ as a function of the physical thicknesses as well as the refractive indices of the respective layers. As the individual layer thicknesses have been fixed via cross sectional imaging, this leaves the refractive indices as free parameters. As this method retrieves the refractive index over a broad bandwidth, dispersion needs to be taken into account.

We choose an adequate dispersion model, based on the wavelength region and the material type.
In general, we found that models with fewer parameters, such as the model developed by Afromowitz~\cite{Afromowitz1974} used in our experimental study, are more robust against overfitting to remaining systematic measurement deviations (for details see~\cite{PernerSimultaneously-MeasuredGaAs}).
However, the model needs to be able to capture expected dispersion variations adequately. Depending on the situation, this criterion can be satisfied by (modified) Sellmeier-type empirical models~\cite{Sellmeier1871ZurSubstanzen, Gehrsitz2000TheModeling}, various semi-empirical models~\cite{Afromowitz1974, Wemple1971BehaviorMaterials} and models derived from first principles~\cite{Forouhi1988OpticalDielectrics}.

Finally, running a non-linear least-squares routine to fit the TMM model to the transmittance data, based on measured layer thicknesses and the chosen refractive index model, provides best fit parameters (see Fig.~\ref{fig:T_spectrum_fit_and_n_results}(a)). We use these parameters to pin the refractive indices models over a broad wavelength range (see Fig.~\ref{fig:T_spectrum_fit_and_n_results}(b--c)).
As both the layer thicknesses, which seed the model, and the transmittance measurements, constituting the data that the model is fit to, have an associated measurement uncertainty, error propagation to the best fit parameters is not straightforward and no standard procedure exists.

To overcome this challenge, we calculate the propagated uncertainty using a Monte-Carlo-type method. For that, we randomly varied all measurements (layer thicknesses and transmittance values) according to their associated measurement uncertainties and repeatedly calculated the best-fit parameters. This procedure leads to a distribution of different fit parameters, with its standard deviation representing the propagated standard uncertainty.

To demonstrate the feasibility and performance of this simultaneous refractive index measurement approach, we probed a MBE-grown crystalline GaAs/\AlGaAsx{} multilayer structure~\cite{PernerSimultaneously-MeasuredGaAs}. Specimens previously produced by the same manufacturer under identical conditions were recently shown to have ultra-low absorption and scatter losses~\cite{Winkler2021Mid-infraredPpm, Truong2022Transmission-dominated000}, making this material system an ideal sample.
The specimen used in this study was designed as a HR DBR mirror with a design wavelength $\lambda_{\mathrm{d}}$ of 4.5 microns with 22.5 pairs of AlGaAs/GaAs layers with a nominal optical thickness $d_{i,\mathrm{opt}} = n_id_i$ of $\lambda_{\mathrm{d_i}}/4$ for both materials, where $n_i$ and $d_i$ are the refractive index and physical thickness of the $i^\mathrm{th}$ layer, respectively. That way, reflected waves interfere constructively in a region around $\lambda_{\mathrm{d}}$. This design leads to the aforementioned stopband structure around $\lambda_{\mathrm{d}}$ (see Fig.~\ref{fig:T_spectrum_fit_and_n_results}(a)). The periodic structure was terminated with a single GaAs layer of $\lambda_{\mathrm{d}}/8$ to avoid oxidation of the topmost \AlGaAsx{} layer. Directly after growing the structure, the manufacturer determined the AlAs mole fraction of the \AlGaAsx{} layers to be $x=\SI{0.929(30)}{}$ via X-ray diffraction (XRD).

A photometrically accurate transmittance spectrum was obtained using a commercial Fourier-transform infrared (FTIR) Spectrometer (Bruker Vertex 80v).
We found that the systematical experimental errors were minimized by using a stabilized incandescent SiC globar light source, a KBr beamsplitter, and a pyroelectric DLaTGS detector. Furthermore, the device was carefully aligned and evacuated (to minimize atmospheric absorption in the free-space path) for several hours prior to the measurement series, also letting the light source thermalize to minimize drifts.
For the sample measurements, we aligned the specimen to normal incidence on a standard optics mount and a thermoelectric Peltier-type cooler was attached at the base to maintain a constant sample temperature of \SI{22(1)}{\degreeCelsius}.
Multiple averaged spectra were recorded and statistically evaluated to obtain the type A uncertainty.
Care was also taken in choosing appropriate parameters during the Fourier-transformation of the recorded interferometric data to obtain the final transmittance spectrum in Fig.~\ref{fig:T_spectrum_fit_and_n_results}(a).

To obtain the layer thicknesses via cross-sectional SEM imaging, we cleaved the sample along one of the crystal axes. In order to avoid boundary effects in the topmost layer, a thin gold coating was sputtered on top of the layer structure prior to cleaving (see Fig.~\ref{fig:Intro}).
The measurement accuracy was ensured by calibrating the imaging with a traceable and certified calibration standard (EM-Tec MCS-0.1CF), while the precision was achieved by evaluating the resulting image line-by-line. This was possible due to the excellent resolution and SNR of the backscattered-electron detector as well as the excellent lateral thickness uniformity of the heterostructure, which is typical for such multilayer structures~\cite{Koch2019ThicknessInterferometry}.
The layer boundaries in each line were found by fitting a function to each step-like feature, to subsequently deduce the layer thicknesses.
These per-line measurements of a total of four images were then averaged to drastically reduce the statistical uncertainty. This way, the only significant error source was the calibration, which was propagated from the calibration routine to the individual layer thicknesses (see Fig.~\ref{fig:SEM_overview}(b)).

To model the wavelength dependence of the refractive indices in the best-fit routine, we chose the model developed by Afromowitz~\cite{Afromowitz1974} for both materials. To enable this approach, the band gap energies are taken from literature~\cite{Sell1974ConcentrationEV, Aspnes1986OpticalAlxGa1-xAs}. Furthermore, we used literature values from~\cite{Skauli2003ImprovedOptics} to model the GaAs seed wafer. This demonstrates that the substrate does not need to match one of the layer materials. The TMM model was calculated by a implementation of the TMM-Fast Python package~\cite{Luce2022TMM-FastTutorial}.

As can be seen from Fig.~\ref{fig:T_spectrum_fit_and_n_results}(a), the best-fit model closely follows the features of the measured transmittance spectrum, even where the features deviate from the spectrum expected from a strictly periodic structure. Of special note is the feature at \SI{\sim 2.2}{\um}, where the systematic deviation of the DBR layer's thicknesses from the respective mean values (as measured for layers No. 40 and up in Fig.~\ref{fig:SEM_overview}(a)) lead to an irregularity in the spectrum. The deviation at \SI{\sim 4.7}{\um}, on the other hand, is best explained by a remaining systematic error on the thickness of the $\lambda/8$ cap, as we could show by varying the layer after obtaining best-fit values.

Finally, we obtained uncertainties to the retrieved refractive index results via a Monte-Carlo-type uncertainty propagation routine.
This involves running the fit routine 1000 times, where each time the transmittance values, the individual layer thicknesses and literature values are randomly picked from distributions representing their respective measurement uncertainty. For each run, we fit the free parameters and calculate both refractive indices. Repeating this procedure results in distributions for $n(\lambda)$, from which we infer the mean and uncertainty of the final results, as shown in Fig.~\ref{fig:T_spectrum_fit_and_n_results}(b--c). The resulting relative standard uncertainty is $s(n_{\mathrm{GaAs}})/n_{\mathrm{GaAs}} \leq \SI{3.14e-4}{}$ and $s(n_{\mathrm{AlGaAs}})/n_{\mathrm{AlGaAs}} \leq \SI{2.80e-4}{}$ for the respective materials.

In summary, we detail a general, highly adaptable method to simultaneously measure the refractive indices of two materials simultaneously by probing a thin-film multilayer heterostructure, ideally a simple quarter-wave DBR.
The individual steps involve acquiring a transmittance spectrum, as well as an accurate measurement of the layer thicknesses. This is followed by a non-linear least-squares fitting routine of a TMM model with the refractive indices modelled according to a suitable empirical, semi-empirical or theoretical model. The best fit results are then used to obtain the refractive indices over a broad wavelength range.
Propagation of the measured uncertainties to the the model's best fit parameters is realized via a Monte-Carlo-type routine. This approach is robust to common systematic measurement errors, as the refractive indices are tightly constrained by the measured quantities.

The feasibility of the method is verified by an experimental study on a GaAs/\AlGaAsx{} multilayer~\cite{PernerSimultaneously-MeasuredGaAs}, which yields results in good agreement with previously-published results and propagated uncertainties on the \SI{e-4}{} level in the \SIrange[range-units = single]{2}{7}{\um} spectral range. Here, an FTIR spectrometer is used to obtain the photometrically accurate transmittance spectrum and the accurate layer thicknesses are obtained via calibrated SEM metrology. Careful control of systematic uncertainties was necessary but achieved by simple means such as proper alignment and temperature stabilization.

Compared to other approaches, the presented routine realizes high levels of accuracy and precision, while drastically reducing experimental complexity. The transmittance spectrum can be of low resolution ($>\SI{2}{\per\cm}$), as long as the broadband spectral features of DBR-type multilayers are adequately resolved.
We therefore do not require separate, specialized, and cost intensive optical setups, such as a spectroscopic ellipsometers~\cite{Fujiwara2007SpectroscopicEllipsometry}, instead relying on devices commonly available in optics laboratories.
In the evaluation step, we avoid intricate extrapolation routines, which are needed in the fringe pattern analysis used for FTIR refractometry~\cite{Kaplan1998FourierRefractometry, Skauli2003ImprovedOptics}.

The employed multilayer samples are widely available in many different materials, making this method an ideal candidate for routine measurements.
Extending the proposed method to amorphous dielectric multilayers will allow to cover most relevant optical materials. These multilayers bear the advantage that one of the materials is not exposed to atmosphere, easing the measurement of the refractive index of materials that would quickly form oxide layers when exposed to air.

This method will allow to routinely and accurately measure refractive indices of materials in their transparent range. This is of special importance for the mid-infrared range, which is of high interest for applications in spectroscopy, but also a region where the optical properties of many materials are still poorly studied.

\begin{acknowledgments}
We thank Dr. Valentina Shumakova for her help with creating the infographic in Fig. \ref{fig:Intro}.

We acknowledge support by the Faculty Center for Nano Structure Research at the University of Vienna, providing the SEM. The financial support by the Austrian Federal Ministry for Digital and Economic Affairs, the National Foundation for Research, Technology and Development and the Christian Doppler Research Association is gratefully acknowledged.
\end{acknowledgments}

\bibliography{references-final}

\end{document}